\documentclass{iau} 
\usepackage{graphicx}
\usepackage{hyperref}
\usepackage{afterpage}

\usepackage[dvipsnames]{xcolor}
\usepackage{natbib}
\definecolor{linkcolor}{rgb}{0.0,0.3,0.5}
\definecolor{urlcolor}{rgb}{0.27,0.55,0.}
\definecolor{funcolor}{rgb}{0.65, 0.16, 0.16}
\hypersetup{colorlinks=true,linkcolor=linkcolor,citecolor=linkcolor,filecolor=linkcolor,urlcolor=linkcolor}

\title[Black hole priors] %
{Reanalysis of LIGO black-hole coalescences with alternative prior assumptions}

\author[Davide Gerosa \emph{et al.}]   %
{Davide Gerosa$^1$
\thanks{Einstein Fellow}, Salvatore Vitale$^2$, Carl-Johan Haster$^3$, \\Katerina Chatziioannou$^3$
\and Aaron Zimmerman$^3$}

\affiliation{$^1$TAPIR 350-17, California Institute of Technology, \\1200 E California Boulevard, Pasadena, CA 91125, USA \\ 
$^2$LIGO, Massachusetts Institute of Technology, \\ 77 Massachusetts Avenue, Cambridge, MA 02139, USA\\ 
$^3$Canadian Institute for Theoretical Astrophysics, University of Toronto, \\ 60 St. George Street, Toronto, ON M5S 3H8, Canada\\
{ \href{mailto:dgerosa@caltech.edu}{dgerosa@caltech.edu}} }

\pubyear{2017}
\volume{338}  %
\jname{Gravitational Wave Astrophysics}%
\editors{Editor: G. Gonz\'alez}
\begin{document}

\maketitle

\begin{abstract}
We present a critical reanalysis of the black-hole binary coalescences detected during LIGO's first observing run under different Bayesian prior assumptions. We summarize the main findings of \cite{2017arXiv170704637V} and show additional marginalized posterior distributions for some of the binaries' intrinsic parameters.
\vspace{0.1cm}\\
These findings were presented at IAU Symposium 338, held on 
October 16-19, 2017 in Baton Rouge, LA, USA. 

\end{abstract}
\firstsection %
\section{The incredible story of the first LIGO and Virgo events}

The first two observing runs of the advanced gravitational wave (GW) detectors LIGO and Virgo \citep{2015CQGra..32g4001L,2015CQGra..32b4001A} are an incredible story of scientific achievement (and possibly a pinch of luck).

The first observing run (O1) lasted from September 2015 to January 2016, during which the two LIGO detectors in Hanford, WA and Livingston, LA collected about $\sim 51.5$ days hours of coincident data. Three detections were made during this first data taking period, all from black hole (BH) binary systems \citep{2016PhRvX...6d1015A}. A few days \emph{before} the official start of the scientific run, the strong event GW150914 from a BH binary with total mass $\sim 65 M_\odot$ hit the two LIGO detectors with a signal-to-noise ratio (SNR) of $\sim 24$ \citep{2016PhRvL.116f1102A}. This first landmark observation of GWs is one of the greatest achievements in modern science and crowned with success more than $50$ years of experimental and theoretical effort in building and operating GW detectors. A few months later, on a Christmas day evening in the US, LIGO detected a second signal from a lighter BH binary of $\sim 24 M_\odot$, GW151226 \citep{2016PhRvL.116x1103A}. A weaker and less significant event, LVT151012, was also detected during O1.  The probability that LVT151012 was not instrumental noise is 87\% \citep{2016PhRvX...6d1015A}. It turns out this is not enough to qualify for the ``GW'' stamp, and the event was designated to be a LIGO/Virgo trigger.

The second observing run (O2) lasted from November 2016 to August 2017. The earliest two detections announced during O2 curiously mirrored the O1 results. First, GW17014 was detected from a BH binary of $\sim 50 M_\odot$ similar to that of GW150914 \citep{2017PhRvL.118v1101A}. Later on, GW170608 was detected from a BH binary of $\sim 19 M_\odot$ which resembles GW151226 \citep{2017arXiv171105578T}. The real surprises came towards the end of the data taking period. GW170814 was the first event detected simultaneously by the two LIGO detectors in the US and the Virgo interferometer in Europe \citep{2017PhRvL.119n1101A}. Adding the third detector to the network allowed for a drastic reduction of the sky location error box (from $\sim 1000\,{\rm deg}^2$ to $\sim 100\,{\rm deg}^2$), opening  the possibility of rapid electromagnetic follow-up campaigns. In a remarkable twist of events, this possibility was realized only three days later, when GW170817 hit the LIGO detectors and was not confidently detected by Virgo \citep{
2017PhRvL.119p1101A}. Virgo however, was taking data at the time, meaning the signal came from close to the blind spot of the interferometer. This allowed for an incredibly accurate sky localization within $\sim 27\,{\rm deg}^2$. The estimated masses of GW170817 are compatible with those of neutron stars, whose merger is expected to produce a variety of electromagnetic signatures. Indeed, the same event was seen in gamma rays just $\sim 1.7$ s after merger and, in just a few hours, several observatories identified an optical transient in  NGC 4993, a lenticular galaxy at a mere distance of $\sim 40$ Mpc \citep{2017ApJ...848L..12A}. This triggered an extensive follow-up campaign in all electromagnetic bands, providing us unique insights on neutron star mergers, short gamma-ray bursts, and kilonovae.

The parameters of all LIGO/Virgo events were estimated using powerful statistical pipelines which inevitably include prior assumptions. Here we present a critical reanalysis of the three O1 detections (GW150914, GW151226, and LVIT151012) under different Bayesian priors, summarizing results presented by
\cite{2017arXiv170704637V}. We repeat some references and context from \cite{2017arXiv170704637V} but refer the reader to that work for full details.
This was the first independent reanalysis of the public LIGO data that made astrophysical inferences about the sources of the signals. This study pioneered the use
of the scientific products released by the LIGO and Virgo Collaboration 
\citep[\href{https://losc.ligo.org}{losc.ligo.org},][]{2015JPhCS.610a2021V} to uncover finer details of these landmark discoveries. %

\section{Bayesian statistics: a magnifying lens for experimental data}
\label{bayessec}

Current BH binary data were analyzed using Bayesian statistics. In a nutshell, Bayesian statistics aims at improving our understanding of any given phenomenon (like BH mergers) by updating previous knowledge in light of new data. %
This principle is encoded in Bayes' theorem
\begin{equation}
P(\theta | d) =\frac{ P(d|\theta) p(\theta) }{\int P(d| \theta) p(\theta) d\theta}\;.
\label{bayes}
\end{equation}
The posterior probability $P(\theta|d)$ of measuring parameters $\theta$ in light of some new data $d$ is proportional to both the likelihood $P(d|\theta)$ of measuring $d$ and our prior beliefs $p(\theta)$. 
Much like a magnifying lens allows to spot finer details, Bayesian statistics is a natural procedure to make the information brought by the data evident against the background of our prior knowledge. Crucially, such background needs to be specified.

In GW research, we all carry a monumental prior, namely the theory of General Relativity (GR). GR, indeed, has passed all current experimental tests with flying colors. 
It is therefore reasonable to assume that GR is an accurate description of reality, even if it may not be the ultimate theory of gravity.
This does not mean that GR is not put to the test with GW data, but rather that it is more reasonable to attempt measurements of \emph{deviations from GR}, rather than measuring the absolute theory of gravity. This approach just reflects previous experience, corroborated by $\sim 100$ years of data, that saw GR coming out of any experimental test stronger than before.

\section{What prior knowledge \emph{could} go into a black hole analysis?}

Astrophysical BHs (in GR at least) are fully characterized by two quantities, namely their mass and spin. A BH binary is therefore described by eight intrinsic parameters $\theta=\{m_1,m_2,\boldsymbol{\chi_1},\boldsymbol{\chi_2}\}$.
Let us describe the direction of each spin vector $\boldsymbol{\chi_i}$ with a polar angle $\theta_i$ measured from the binary's orbital angular momentum $\mathbf{L}$ and an azimuthal angle $\phi_i$ measured in the orbital plane. The relative orientations of the spins and the orbital angular momentum depend on the three angles $\theta_1,\theta_2$ and $\Delta\Phi=\phi_2-\phi_1$.

Prior distributions have to be specified on all these parameters when analyzing BH coalescence data. All LIGO/Virgo analysis  were performed with uniform prior distributions in $(m_1,m_2),\chi_1,\chi_2, \cos\theta_1,\cos\theta_2,\phi_1$ and $\phi_2$ at the reference frequency of $f_{\rm ref}=20$ Hz \citep{2016PhRvX...6d1015A,2016PhRvL.116x1103A,2016PhRvL.116f1102A,2017PhRvL.118v1101A,2017arXiv171105578T,2017PhRvL.119n1101A,2017PhRvL.119p1101A}. This is a very  reasonable approach: we have never observed BHs in binaries before, so there is no reason to prefer a particular mass or spin range. However, this is surely not the only reasonable approach. For instance, spins are vectors, not scalar. One might prefer to assume spin vectors uniform in volume, rather than uniform in magnitude and isotropic in direction. We also have solid understanding that massive stars end their lives as BHs, following core collapse. Should we insert this piece of information into our BH analysis? The distribution of stellar masses follow a power-law distribution $p(m)\propto m^{\alpha}$ with $\alpha\sim -2.3$. Could this be a good candidate for a BH mass prior?

\cite{2017arXiv170704637V} first tried to address some of these issues by fully reanalyzing the O1 LIGO data with different prior assumptions. Their prior choices are summarized as follows:
\begin{itemize}
\item[($P_1$)] Masses are uniform, spin magnitudes are uniform,  spin directions are isotropic. This is the standard prior choice made by the LIGO/Virgo Collaboration.
\item[($P_2$)] Masses are uniform, spin magnitudes are uniform in rotational energy, spin directions are isotropic.
\item[($P_3$)] Masses are uniform, spin magnitudes are uniform in volume, spin directions are isotropic.
\item[($P_4$)] Masses are uniform, spin magnitudes are drawn from a bimodal distribution peaked at low and large spins, spin directions are isotropic.
\item[($P_5$)] Masses are uniform, spin magnitudes are uniform, spin directions are preferentially aligned with $\mathbf{L}$ (i.e. small $\theta_i$).
\item[($P_6$)] Masses of the primary BHs are drawn from the stellar initial-mass function, secondary masses are uniform, spin magnitudes are uniform and spin directions are isotropic;
\item[($P_7$)] Masses of the primary BHs are drawn from the stellar initial-mass function, mass ratio is drawn from a logistic distribution, spin magnitudes are uniform and spin directions are isotropic.
\item[($P_8$)]  Masses are uniform, spins magnitudes are preferentially low, spins directions are isotropic.
\end{itemize}

As detailed in \cite{2017arXiv170704637V}, these choices leverage some know astrophysical properties of stellar-mass BHs and/or theoretical predictions on their population in binaries. 

 \begin{figure}[p]
\includegraphics[page=1,width=\textwidth]{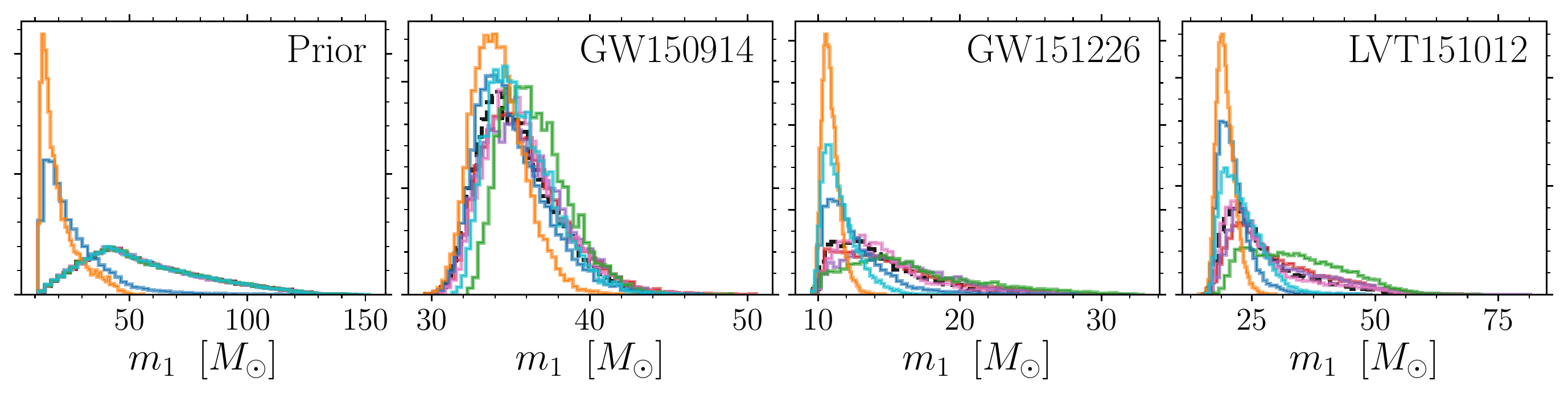}
\includegraphics[page=2,width=\textwidth]{procfig}
\includegraphics[page=3,width=\textwidth]{procfig}
\includegraphics[page=4,width=\textwidth]{procfig}
\includegraphics[page=5,width=\textwidth]{procfig}
\includegraphics[page=8,width=\textwidth]{procfig}
\caption{Prior (left panels) and posterior densities  for GW150914, GW151226 and LVT151012 (second, third and fourth column from the left) under a variety of prior assumptions $P_1$-$P_8$. We show marginalized distributions for the BH component masses $m_1$ and $m_2$, total mass $M$, mass ratio $q$, chirp mass $M_c$ and effective spin $\chi_{\rm eff}$. 
The $\chi_{\rm eff}$ panels were already presented by \cite{2017arXiv170704637V}.}
\label{marg1}
\end{figure}

\begin{figure}[p]
\includegraphics[page=6,width=\textwidth]{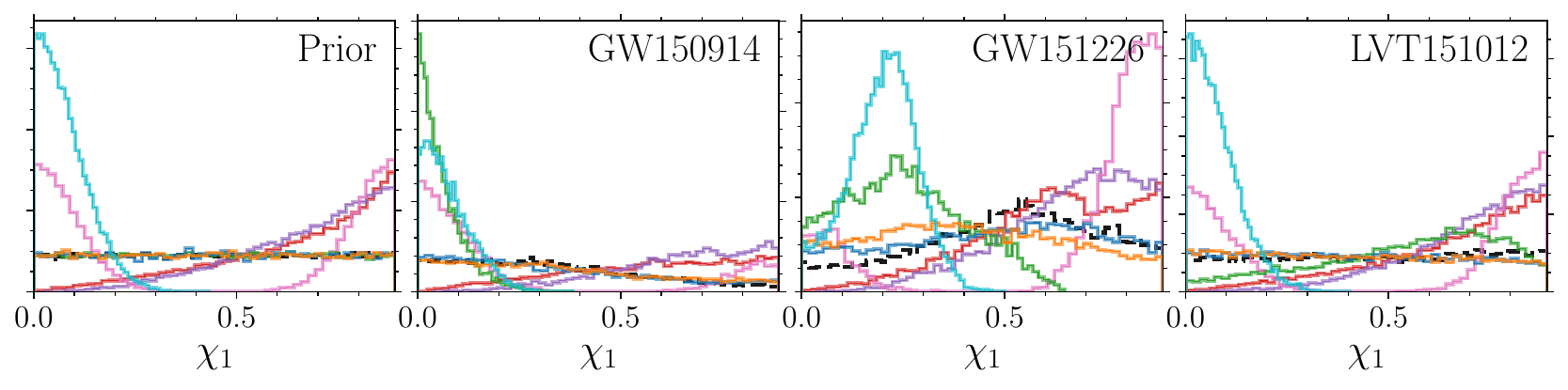}
\includegraphics[page=7,width=\textwidth]{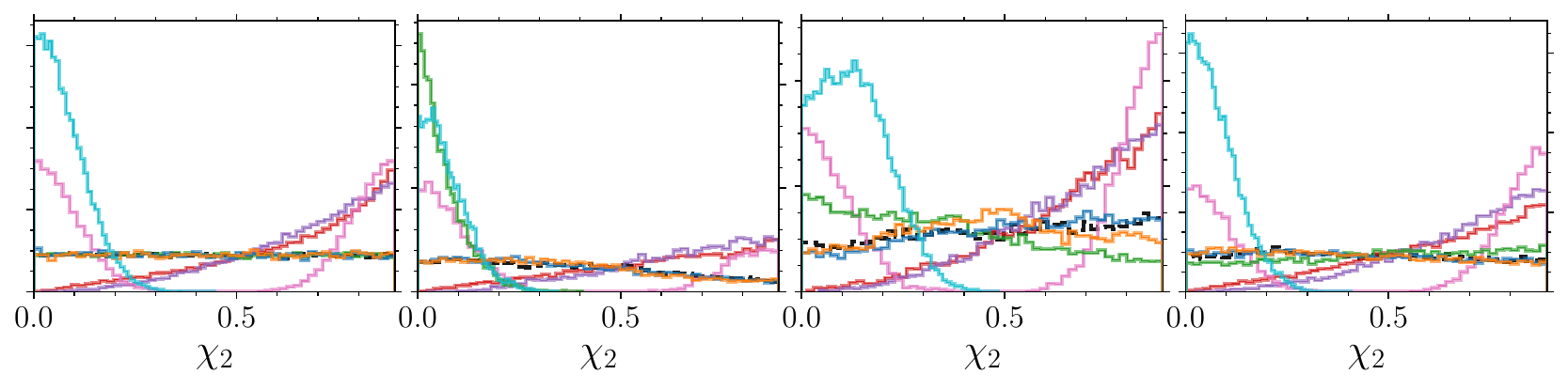}
\includegraphics[page=10,width=\textwidth]{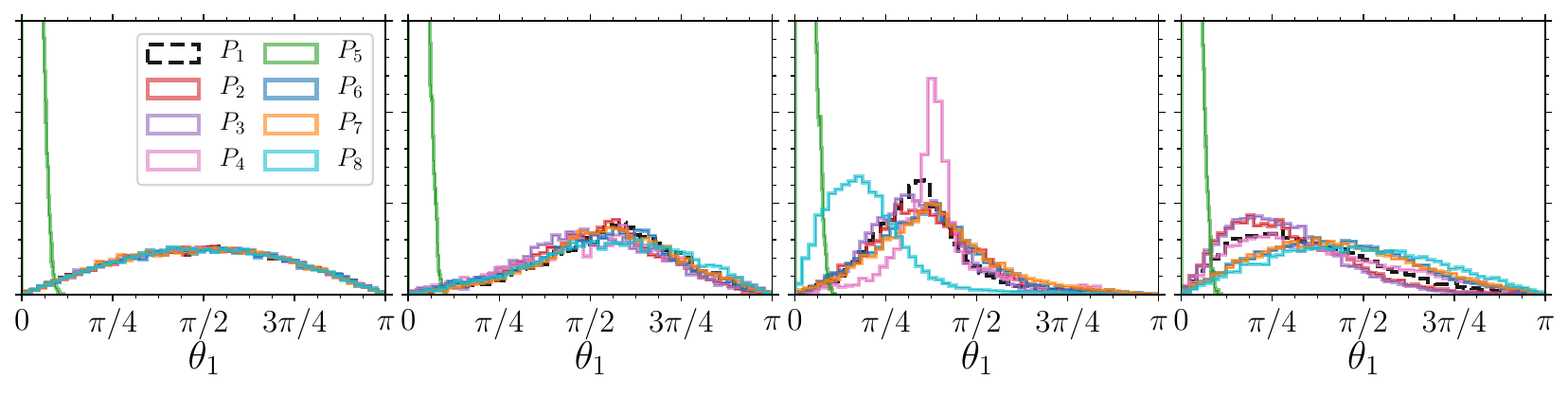}
\includegraphics[page=11,width=\textwidth]{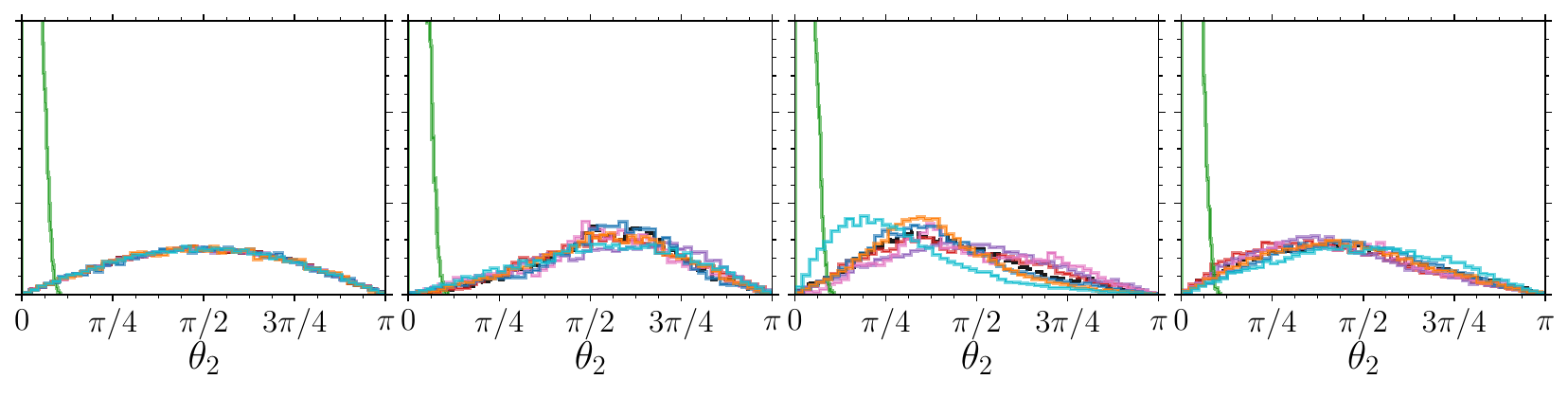}
\includegraphics[page=12,width=\textwidth]{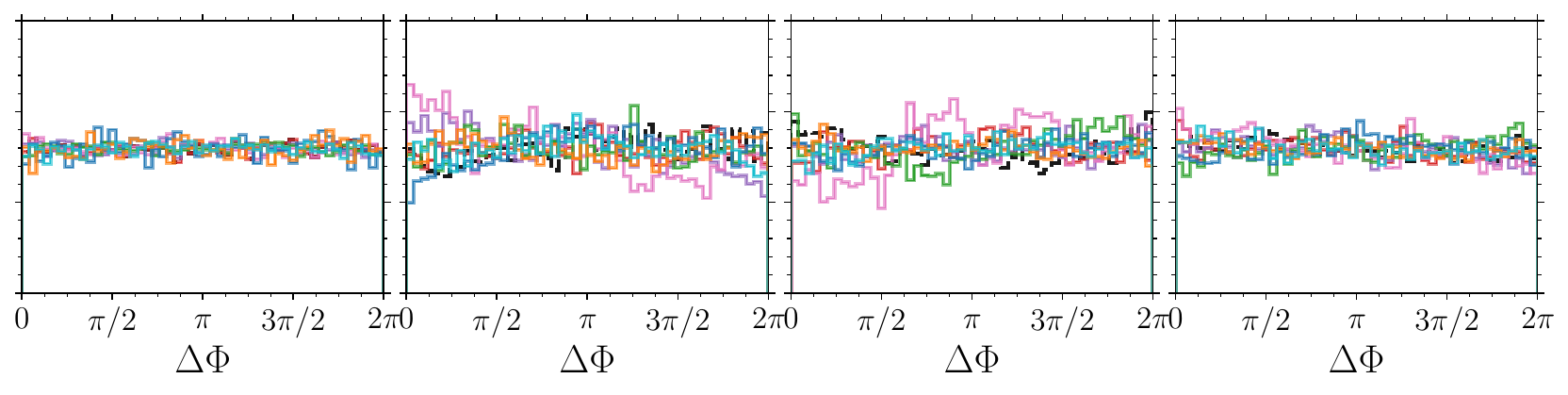}
\includegraphics[page=9,width=\textwidth]{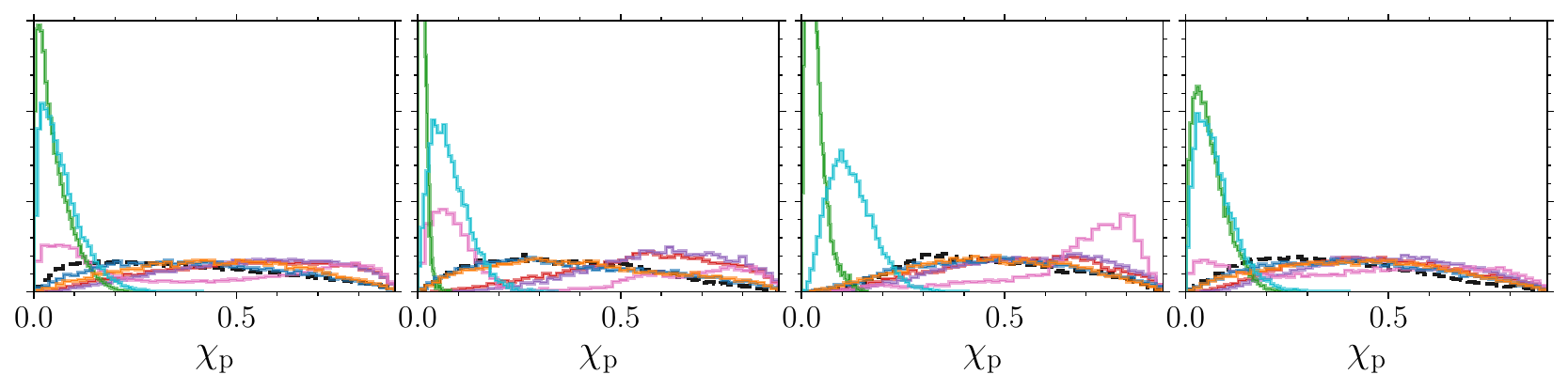}
\caption{Prior (left panels) and posterior densities for GW150914, GW151226 and LVT151012 (second, third and fourth column from the left) under a variety of prior assumptions $P_1$-$P_8$. We show marginalized distributions for the BH spin magnitudes $\chi_1$ and $\chi_2$, the tilt angles $\theta_1$ and $\theta_2$ measured from the orbital angular momentum, the angle between the projections of the two spins onto the orbital plane $\Delta\Phi=\phi_2-\phi_1$, and the precession parameter $\chi_p$. Spin angles are reported at the reference frequency $f_{\rm ref}=20$ Hz.}
\label{marg2}
\end{figure}

\section{Marginalized posteriors under different priors}
\label{marginalized}

In Figs.~\ref{marg1} and \ref{marg2} we shows marginalized posterior distributions for a variety of intrinsic parameters that characterize the observed BH events and all our prior choices $P_1$-$P_8$. Details on the data analysis technique are reported by \cite{2017arXiv170704637V}. The two component masses $m_1$ and $m_2$ can be combined into total mass $M=m_1+m_2$, mass ratio $q=m_2/m_1\leq 1$ and chirp mass $M_c=   (m_1 m_2)^{3/5}/M^{1/5}$. The best measured spin quantity is the effective spin $\chi_{\rm eff} = (m_1 \chi_1\cos\theta_1+m_2\chi_2\cos\theta_2)/M$; spin precession is implemented into the waveform template used in this analysis through a single parameter $\chi_p$. %

Some of the inferred physical properties are robust under the choice of the Bayesian priors. For instance:
\begin{enumerate}
\item non-spinning BHs for GW151226 are excluded at $>90\%$ credible interval;
\item conversely, GW150914 and LVT151012 can be described by $\chi_1=\chi_2=0$;
\item mass ratios $q\lesssim 0.5$ for GW150914 are excluded at $>90\%$ credible interval;
\item the 90\% credible intervals in $M_c$ change by $<1M_\odot$ in all cases;
\end{enumerate}
On the other hand, other parameters strongly depend on the prior. Notably, this includes component masses $m_i$ and spin magnitudes $\chi_i$.
The spin angles $\theta_1$, $\theta_2$, $\Delta\Phi$ and the spin precession parameter $\chi_p$ cannot be constrained meaningfully, under any of the priors tested here. We note, however, that this could be due to the waveform approximant used by \cite{2017arXiv170704637V} which only implements precession through a single effective spin. %
 In general, prior effects are more severe for low SNR events like LVT151012, where data are less informative. A thorough discussion of these results (including odds ratio calculations) has been presented by  \cite{2017arXiv170704637V}. We encourage the reader to compare the findings spelled out by \cite{2017arXiv170704637V} with Fig.~\ref{marg1} and \ref{marg2} of these proceedings.

\section{Was it necessary?}

Once a posterior $P(\theta|d)$ has been obtained from a reference prior $p(\theta)$ (say $P_1$ from above), one can in principle use Bayes theorem (\ref{bayes}) to obtain the posterior $\tilde P(\theta|d)$ for a different prior $\tilde p(\theta)$:
\begin{equation}
\tilde P(\theta|d) \propto  P(\theta|d) \frac{\tilde p(\theta)}{p(\theta)}\;.
\label{reweight}
\end{equation}
This prompts the following question: was it necessary to fully reanalyze the data to obtain the posteriors presented in Sec.~\ref{marginalized}? Couldn't one just re-weight previous posterior samples as in Eq.~(\ref{reweight})? While this procedure is mathematically correct, it might be dangerous in practice, because the numerical algorithm used to sample $P$ might not cover the region of the parameter space that is of interest for $\tilde P$ densely enough. This issue was recently explored by \cite{2017arXiv170903095W}: they find that re-weighted posterior samples might indeed lead to biassed conclusions. Re-weighting carries systematics which needs to be properly analyzed before Eq.~(\ref{reweight}) can be used in, e.g., hierarchical model selection schemes. Careful comparisons of our full reruns with re-weighted posteriors is a natural follow-up of the results presented by \cite{2017arXiv170704637V} and summarized here.

 \acknowledgements 
D.G. is supported by NASA through Einstein Postdoctoral Fellowship
Grant No. PF6-170152 by the Chandra X-ray Center,
operated by the Smithsonian Astrophysical Observatory
for NASA under Contract NAS8-03060.

\bibliographystyle{yahapj}
\bibliography{iauproc}

\end{document}